\documentclass[12pt,a4paper]{article}
\usepackage[margin=2cm]{geometry}
\usepackage{authblk}

\usepackage[breaklinks]{hyperref}
\usepackage{booktabs}
\usepackage{txfonts}

\usepackage{epsfig}
\usepackage{braket}
\usepackage{graphicx}

\usepackage{amsthm}
\usepackage{amsmath}
\usepackage{amssymb}
\usepackage{amsfonts}
\usepackage{mathbbol}
\usepackage{caption}
\usepackage{bm}
\usepackage{xcolor}
\usepackage{mathrsfs}
\usepackage{wrapfig}

\newcommand{\be}{\begin{equation}}
\newcommand{\ee}{\end{equation}}
\newcommand{\bea}{\begin{eqnarray}}
\newcommand{\eea}{\end{eqnarray}}
\def\bml{\begin{subequations}}
	\def\blea{\bml\begin{eqnarray}}
	\def\eml{\end{subequations}}
	\def\elea{\end{eqnarray}\eml}

\begin{document} 
	
	\title{Modeling the microlensing of circumbinary systems}
	
	\author[1]{Brett George}
	\author[1,2,3]{Eleni-Alexandra Kontou\thanks{\tt e.a.kontou@uva.nl}}
	\author[1]{Patrycja Przewoznik }
	\author[2]{Eleanor Turrell}
	\affil[1]{Department of Physics, College of the Holy Cross, Worcester, Massachusetts 01610, USA}
		\affil[2]{Physics Program, Bard College, 30 Campus Rd, Annandale-on-Hudson, NY 12504, USA}
	\affil[3]{ITFA and GRAPPA, Universiteit van Amsterdam, Science Park 904, Amsterdam, the Netherlands
	}
	\date{\small\today}
	\maketitle

	\begin{abstract}
	Gravitational microlensing is one of the methods to detect exoplanets; planets outside our solar system. Here we focus on theoretical modeling of three lens systems and in particular circumbinary systems. Circumbinary systems include two stars and a planet and are estimated to make up a sizable portion of all exoplanets. Extending a method developed for binary lenses to the three lens case, we explore the parameter space of circumbinary systems producing exact magnification maps and light curves. 
	\end{abstract}

\section{Introduction}

Gravitational lensing was one of the major predictions of the general theory of relativity. It refers to the bending of light from a background source by a massive object between the source and the observer. Gravitational microlensing is a special case of gravitational lensing where multiple images are created with very small separations. Typically, a separation of less than a few milliarcseconds remains unresolved with current capabilities.

Gravitational microlensing is more difficult to observe than strong lensing where the images can be separated. It was not until 1986 when Paczynski \cite{paczynski1986gravitational} proposed microlensing observations in the Magellanic clouds. Since the initial searches, thousand of microlensing events have been observed (see \cite{Gaudi:2010nk} for more references).

Mao and Paczynski \cite{mao1991gravitational} were the first to suggest that gravitational microlensing can  be used to discover exoplanets; planets outside our solar system. The main idea is the following: as a background star passes behind, and close to the line of sight of a foreground star, we have a gravitational lensing event. This is a microlensing event as typically the images are unresolved, creating only a surge in the flux of the source star. If the foreground star hosts planets, that will create an additional perturbation which can be observed.

The first exoplanets were discovered in the 1990's \cite{wolszczan1992planetary, mayor1995jupiter} and in the last few decades thousands of exoplanets have been observed using different methods. The most popular is the transit method, where as a planet passes in front of a star, a dip in its flux is observable. Other methods include Doppler spectroscopy, and more recently direct imaging. 

The first detection of an exoplanet using microlensing was in 2004 \cite{bond2004ogle}. Since then over a hundred exoplanets have been detected. Searching for exoplanets using microlensing presents significant difficulties as planetary deviations in light curves are short lived, not repeatable and can easily be missed. Additionally, planetary parameters such as the mass and orbit, depend on the properties of the host star, which are typically unknown. Thus the most successful observations usually include combination of microlensing followed by high-resolution imaging.  

Despite the difficulties, microlensing offers some advantages compared to the more popular transit and Doppler methods. The most important one is that microlensing is the most sensitive method for small planets further from their host star. In particular planets beyond the so-called ``snow line'' where ice is typically formed, are usually undetectable using other methods. Thus microlensing can provide an important addition to the other methods in an effort to create a representative map of exoplanets in our galaxy and beyond. 

The vast majority of exoplanets discovered orbit a single star. This is merely a artifact of our detection methods as it is estimated that as many as 50\% of the stars in our galaxy are part of binary or multiple star systems. A planet that orbits two stars is called a circumbinary planet and the system a circumbinary system. The first such exoplanet was discovered using pulsar timing \cite{thorsett1993psr}. About 20 circumbinary planets have so far been discovered using different methods. The first circumbinary planet discovered using microlensing was found in 2016 \cite{bennett2016first} and it is the only one to date.

Detecting circumbinary planets using microlensing presents both observational and theoretical challenges. While solving a single lens equation is trivial, adding more bodies significantly increases the complexity. Double lens systems have been well studied theoretically (for example \cite{schneider1986two, bozza1999perturbative}). Triple lens systems that include circumbinary systems involve even more parameters (five instead of two) and a much more complex system of equations, making a theoretical exploration challenging. 

A general three lens system is not analytically tractable. One method is to use symmetries that simplify the problem as in \cite{mao1997properties, rhie2003n}. Perturbation theory \cite{bozza1999caustics} is a different approach that has been studied. One star-two (or multiple) planet systems have received the most attention \cite{gaudi1998microlensing, han2001properties} with the planets usually treated in a perturbative way. Some star-planet-moon systems have also been studied \cite{han2002feasibility}. Circumbinary systems have been studied less. Examples include \cite{han2008distinguishing} and \cite{luhn2016caustic} with the latter examining circumbinary caustics for a large range of the parameter space. 

The goal of this paper is to provide an exploration of the parameter space of circumbinary systems using a direct and exact method. Instead of caustics we study  full magnification maps and light curves. We use a combination of an analytical approach developed first by Witt and Mao \cite{witt1995minimum} for binary lenses and a numerical one using Mathematica. The method can be easily used by undergraduate students to generate both magnification maps and light curves. 

We start with a pedagogical introduction to the lens equation in Sec.~\ref{sec:lens} and continue with the basics of magnification in single and $N$-lens systems in Sec.~\ref{sec:magn}. In Sec.~\ref{sec:triple} we analyze the triple lens case, describing the parameter space and our methodology. In Sec.~\ref{sec:parameters} we explore the parameter space, varying the mass and distance of planet, relative mass and distance of stars and angle of the system. In Sec.~\ref{sec:example} we apply our method to the first circumbinary system discovered using microlensing \cite{bennett2016first}. We conclude in Sec.~\ref{sec:conclusions} and discuss possibilities of future work.

\section{The  lens equation}
\label{sec:lens}

In this section we first analyze the geometry of a single lens with spherical symmetry that illustrates the basics of gravitational lensing. We proceed to generalize and derive the $N$-lens equation. Throughout the section we roughly follow \cite{hartle2021gravity} and \cite{Gaudi:2010nk}.

\subsection{The single lens equation}

\begin{figure}[h] 
	\centering
	\includegraphics[scale=0.3]{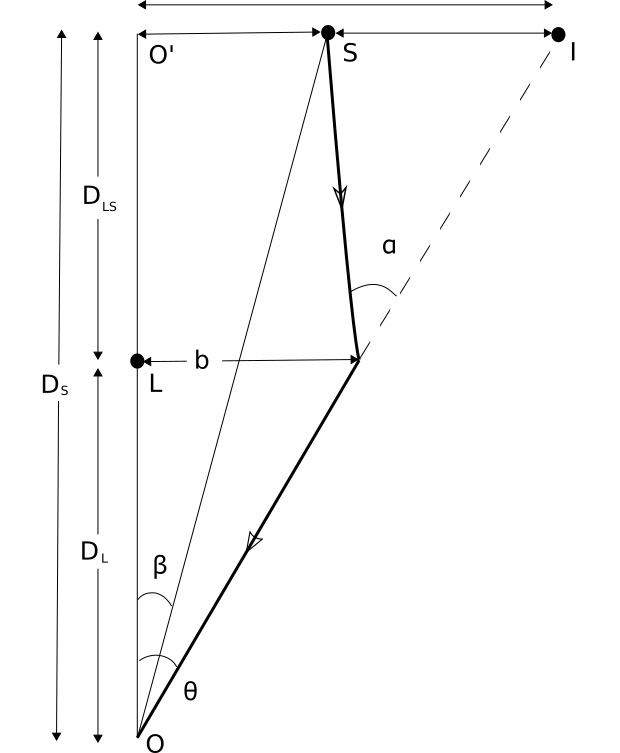}
	\caption{The figure shows the geometry of a single-lens gravitational lensing with spherical symmetry. $O$ is the position of the observer, $S$ the position of the source, $L$ the position of the lens and $I$ the position of the image. $D_L$ is the distance of the lens from the observer, $D_S$ the distance of the source from the observer and $D_{LS}=D_S-D_L$. It is assumed that all the deflection happens transverse to the position of the lens. The path of light is shown in the thick black line and its extension (dashed line) reveals the position of the image. $\beta$ is the angular position of the source and $\theta$ the anglular position of the image, where the angles are measured from the lens. $\alpha$ is the deflection angle given by Eq.~(\ref{eqn:deflection}) and $b$ is the impact parameter. (see also Ref.~\cite{hartle2021gravity}) }
	\label{fig:singlelens}
\end{figure}

The deflection angle for a light ray passing by a mass $M$ at an impact parameter $b$ is given by \footnote{The deflection angle's form can be derived by a purely Newtonian calculation of a massive particle moving with the speed of light, passing by another massive object. The result is the same as Eq.~(\ref{eqn:deflection}) except a factor of two. This factor can only be retrieved by a general relativistic calculation. (see for example Ref.~\cite{hartle2021gravity})}
\be \label{eqn:deflection}
\alpha=\frac{4 R_S}{b}=\frac{4GM}{c^2 b} \,,
\ee
for $b \gg R_S$, where $R_S$ is the Schwarzschild radius. For a spherical lens the geometry of the reflection is given by Fig.~\ref{fig:singlelens}. 

From Fig.~\ref{fig:singlelens}. we have for the transverse distances
\be
O'I=O'S+SI  \,,
\ee
or
\be \label{eqn:tan}
\tan{\theta} D_S=\tan{\beta} D_S+\tan{\alpha} D_{LS} \,.
\ee
In realistic situations all the angles are extremely small so we can use the small angle approximation, $\tan{x} \approx x$, so Eq.~(\ref{eqn:tan}) becomes
\be \label{eqn:lens1}
\theta D_S=\beta D_S+\alpha D_{LS} \,.
\ee
Eq.~(\ref{eqn:lens1}) is called the lens equation. Replacing the deflection angle from Eq.~(\ref{eqn:deflection}) the lens equation becomes
\be \label{eqn:lens2}
\theta=\beta+\frac{4 R_S}{b} \frac{D_{LS}}{D_S} \,.
\ee
From Fig.~\ref{fig:singlelens} the impact parameter $b=\tan{\theta} D_L$ or in the small angle approximation $b=\theta D_L$. So Eq.~(\ref{eqn:lens2}) becomes
\be \label{eqn:lens3}
\theta=\beta+\frac{\theta_E^2}{\theta}  \,,
\ee
where
\be \label{eqn:einstein}
\theta_E=\left[ 2R_S \left(\frac{D_{LS}}{D_S D_L} \right) \right]^{1/2} \,,
\ee
is the Einstein angle. When $\beta=0$ and the source is aligned with the lens, the image position $\theta$ is equal with the Einstein angle. The spherical symmetry implies that in this case the image is a ring around the lens. The ring has physical radius 
\be
R_E \equiv D_L \theta_E \,,
\ee
called the Einstein radius. We can normalize all angles with $\theta_E$  so Eq.~(\ref{eqn:lens2}) becomes 
\be \label{eqn:lensu}
u=y-y^{-1} \,,
\ee
where
\be
u \equiv \frac{\beta}{\theta_E}\,, \qquad y \equiv \frac{\theta}{\theta_E} \,.
\ee
So $u$ and $y$ are the angular positions of the source and the lens respectively in units of the Einstein angle. Solving Eq.~(\ref{eqn:lensu}) gives the positions of the images in terms of the position of the source. Solving the quadratic equation we have for the position of the images
\be \label{eqn:imagepos}
y_{\pm}=\frac{1}{2} (u \pm \sqrt{u^2+4}) \,.
\ee
For $u=0$, there is one image with radius $1$ (Einstein ring). For one of the images $y$ is positive and larger than $1$ while the other is negative and less than $1$. Since the variables are normalized with the Einstein angle that means that one image is always outside the Einstein ring and the other inside as shown in Fig.~\ref{fig:projection}. 

\subsection{The general lens equation}

If we have more than one lenses the geometry it is difficult to visualize the path of the light. However, if we think of the single lens equation (\ref{eqn:lens3}) as a mapping between the angular source position $\beta$ and the angular positions of the images $\theta$, we can easily generalize to the case of $N$ point lenses. The new lens equation is
\be
\bm{\theta}=\bm{\beta}+\bm{\alpha}(\bm{\theta}) \,,
\ee
where $\bm{\alpha}$ is the new deflection angle. The angular positions are now vectors since their directions are not trivial as in the single lens case. The deflection angle for the case of $N$ point lenses of masses $m_i$ is \cite{schneider1984amplification}
\be \label{eqn:lensmultione}
\bm{\alpha}(\bm{\theta})=\frac{4 G D_{LS} }{D_L D_S c^2} \sum_i^N m_i \frac{\bm{\theta}-\bm{\theta}_i}{|\bm{\theta}-\bm{\theta}_i|^2} \,,
\ee
where $\bm{\theta_i}$ are the angular positions of the lenses and we assume that the distances between the lenses are too small compared to the distances of the lenses to the source and the observer. Using the definition of the Einstein angle from Eq.~(\ref{eqn:einstein}) where here $M=\sum_i^N m_i$, we have
\be
\bm{\alpha}(\bm{\theta})=\theta_E^2 \sum_i^N \frac{m_i}{M} \frac{\bm{\theta}-\bm{\theta}_i}{|\bm{\theta}-\bm{\theta}_i|^2} \,.
\ee
Replacing in Eq.~(\ref{eqn:lensmultione}) and defining again the paramerized angular positions of the source, the lenses and the images 
\be
\mathbf{u} \equiv \frac{\bm{\beta}}{\theta_E}\,, \qquad \mathbf{y} \equiv \frac{\bm{\theta}}{\theta_E}\,, \qquad  \mathbf{y}_i  \equiv \frac{\bm{\theta}_i}{\theta_E} \,,
\ee
gives
\be \label{eqn:lensmultitwo}
\mathbf{u}=\mathbf{y}-\sum_i^N \epsilon_i \frac{\mathbf{y}-\mathbf{y}_i}{|\mathbf{y}-\mathbf{y}_i|^2} \,,
\ee
where $\epsilon_i \equiv m_i/M$. 

It is convenient to write the lens equation in complex coordinates. First we define the components of the vectors $\mathbf{u}$, $\mathbf{y}$ and $\mathbf{y}_i$ to be
\be
\mathbf{u}=(u_1,u_2)\, \qquad \mathbf{y}=(y_1,y_2)\,, \qquad \mathbf{y}_i=(y_{i1}, y_{i2}) \,.
\ee
Then we can define complex variables for the source, the images and the lenses positions
\be
\zeta=u_1+i u_2\,, \qquad z=y_1+i y_2 \,, \qquad z_i=y_{i1}+i y_{i2} \,.
\ee
So the lens equation (\ref{eqn:lensmultitwo}) becomes
\be \label{eqn:lenscomplex}
\zeta=z-\sum_i^N \frac{\epsilon_i}{\bar{z}-\bar{z}_i} \,.
\ee

\section{Magnification}
\label{sec:magn}

\subsection{Single lens}

\begin{figure}[h]
	\centering
	\includegraphics[scale=0.3]{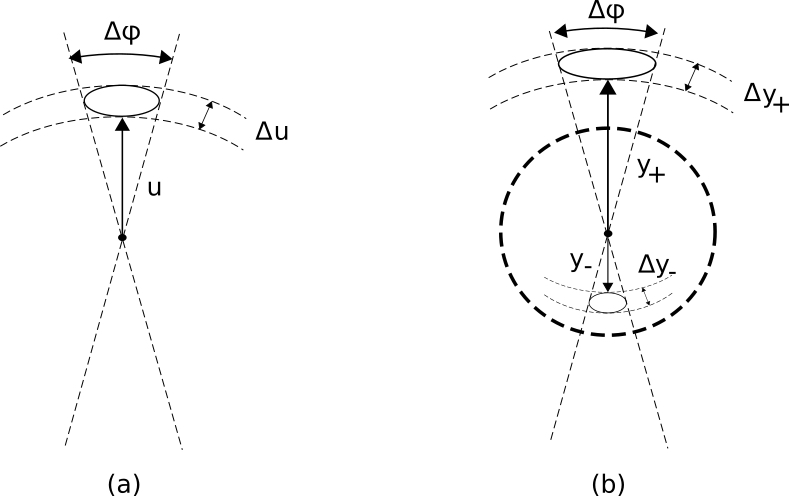}
	\caption{Figure (a) shows the angular size of the unlensed source while figure (b) the two images created with lensing. The dashed circle in figure (b) is the Einstein ring with angular size one according to our parametrization. Image $y_+$ is outside of the ring and larger while image $y_-$ inside of the ring and smaller.}
	\label{fig:projection}
\end{figure}

Lensing preserves the surface brightness of the source but not its flux. The flux is defined as 
\be
\Delta f= (\text{surface brightness}) \times \Delta \Omega \,,
\ee
where $\Delta \Omega$ is the solid angle of the source or the image. 
\be
\Delta \Omega= \sin{\theta} \Delta \theta \Delta \phi \approx \theta \Delta \theta \Delta \phi \,,
\ee
in the small angle approximation. 

The magnification of a image is defined as the flux of the unlensed source over the flux of the image. Following Fig.~\ref{fig:projection} we can write the magnification of one of the images as
\be
A_\pm=\frac{\Delta \Omega_\pm}{\Delta \Omega_*}=\left| \frac{y_\pm \Delta y_\pm}{u \Delta u} \right| \,,
\ee
where the surface brightness cancels because it is conserved, and $\Delta \phi$ cancels since it is the same for source and images. Now since we are looking at the case of a point source we take the limit of $\Delta \theta \to 0$ and get
\be \label{eqn:magnone}
A_\pm=\left| \frac{y_\pm}{u} \frac{d y_\pm}{d u} \right| \,.
\ee
We can take the derivative using Eq.~(\ref{eqn:imagepos})
\be
\frac{d y_\pm}{du}=\frac{1}{2} \left(1 \pm \frac{u}{\sqrt{u^2+4}}\right) \,,
\ee
so 
\be
A_\pm=\frac{1}{2} \left(\frac{u^2+2}{u\sqrt{u^2+4}} \pm 1 \right) \,.
\ee
The total magnification is the sum of the flux of the two images
\be \label{eqn:magnificationone}
A_{\text{tot}}=|A_{+}|+|A_{-}|=\frac{u^2+2}{u\sqrt{u^2+4}} \,.
\ee
As expected, for $u \to \infty$, $A_{\text{tot}} \to 1$, while as $u \to 0$,  $A_{\text{tot}} \to \infty$. 

The source, the lens and the observer are all in motion relative to each other. Here we assume the simplest case; a uniform motion with constant velocity. Since we only care about the relative motion of the system, we assume that only the source is moving and we can parametrize its anglular position $u$ with time. 

\begin{figure}[h]
	\centering
	\includegraphics[scale=0.4]{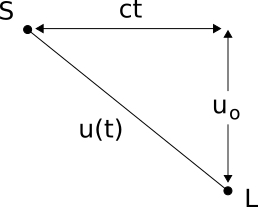}
	\caption{The figure shows the angular position of the source as function of time. $u_0$ is the minimum angular separation between the source and the lens. Since the motion of the source is considered uniform, we can write its angular position in that direction as a linear function of time $ct$. }
	\label{fig:source}
\end{figure}

\begin{figure}[h]
	\centering
	\includegraphics[scale=0.5]{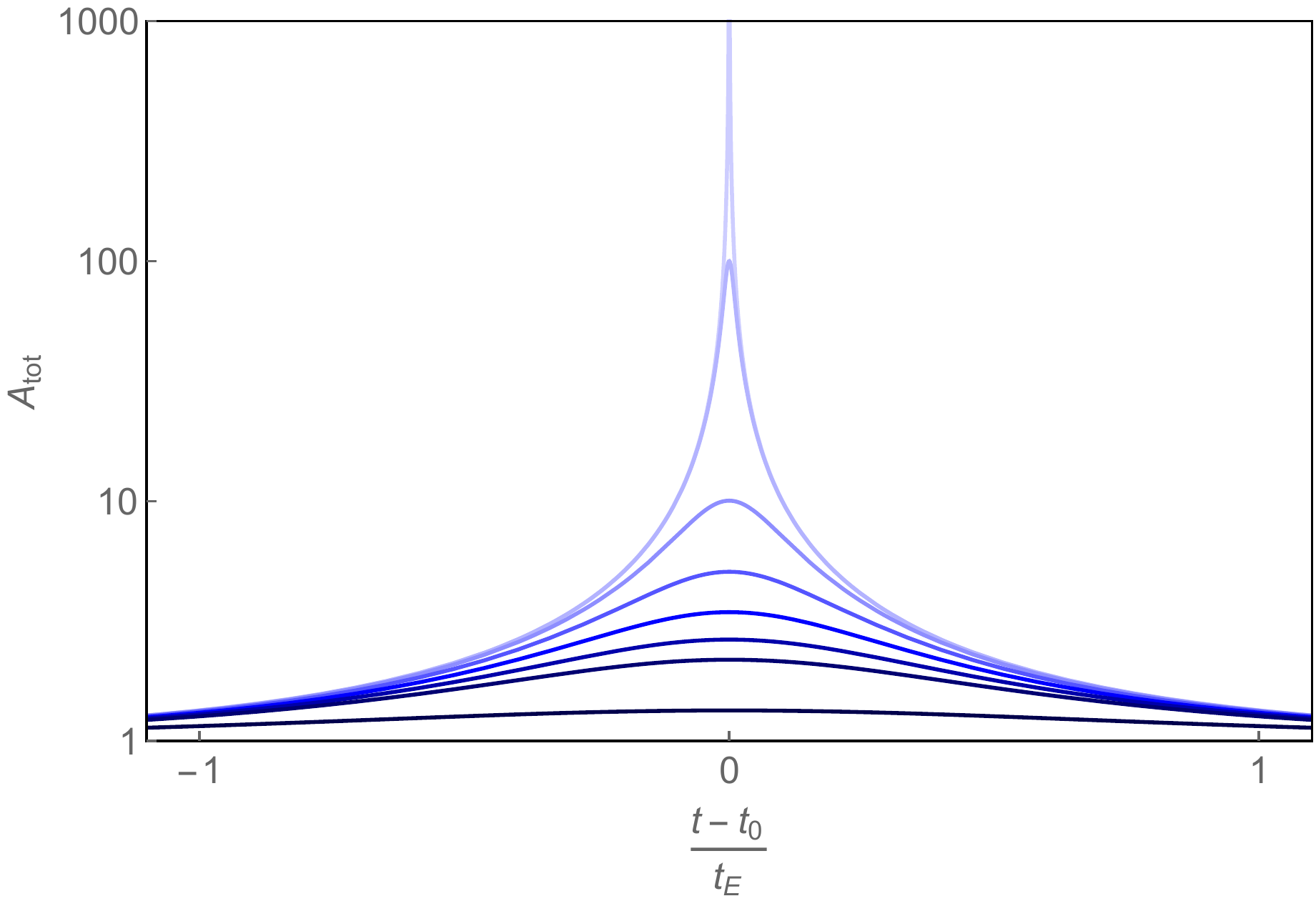}
	\caption{The figure shows light curves for a single lens microlensing event. The curves have $u_0=0, 0.01, 0.1, 0.2, 0.3, 0.4, 0.5, 1$ from lightest to darkest. We can see that for $u_0=0$ the total magnification diverges for $t=t_0$, while for larger $u_0$ the magnification goes to one for all times. }
	\label{fig:single}
\end{figure}

As shown in Fig.~\ref{fig:source} the angular position of the source can be written as
\be \label{eqn:sourceone}
u(t)=\sqrt{u_0^2+(ct)^2} \,,
\ee
where $u_0$ is the minimum angular separation between the lens and the source and $c$ is a constant. Since the angular position of the source is dimensionless and parametrized by the Einstein angle the constant has the inverse units of time $c=1/t_E$ and 
\be
t_E \equiv \frac{\theta_E}{\mu_{\text{rel}}} \,,
\ee
where $\mu_{\text{rel}}$ is the proper motion of the source relative to the lens (see Ref.~\cite{Gaudi:2010nk} for more information). If the minimum angular separation between the source and the lens happens at time $t_0$ and not $t=0$, $t \to t-t_0$ and Eq.~(\ref{eqn:sourceone}) becomes
\be \label{eqn:sourcetwo}
u(t)=\sqrt{u_0^2+\left(\frac{t-t_0}{t_E}\right)^2} \,.
\ee
Using Eq.~(\ref{eqn:magnificationone}) and Eq.~(\ref{eqn:sourcetwo}) we have the total magnification as a function of time, a light curve. Light curves for different values of $u_0$ are shown in Fig.~\ref{fig:single}.

\subsection{General lens}

For a non spherically symmetric lens, as is the case of multiple point masses, a solid-angle element of the source is mapped to a solid angle element of the image \cite{narayan1996lectures}. Then Eq.~(\ref{eqn:magnone}) generalizes to the inverse of the determinant of the Jacobian of the mapping of Eq.~(\ref{eqn:lenscomplex}) and the magnification is given by
\be \label{eqn:magnindi}
A_j=\frac{1}{\det{J}} \bigg|_{z=z_j} \,.
\ee
In our complex notation the Jacobian of the mapping is given by
\be \label{eqn:jac}
\det{J}\equiv \left|\begin{array}{cc}
	\partial u_1/ \partial y_1  & \partial u_1/ \partial y_2 \\
	\partial u_2/ \partial y_1 & \partial u_2/ \partial y_2
\end{array} \right|=\frac{\partial u_1}{\partial y_1}\frac{\partial u_2}{\partial y_2}-\frac{\partial u_1}{\partial y_2}\frac{\partial u_2}{\partial y_1} \,.
\ee
In gravitational lensing $\partial u_1/ \partial y_2=\partial u_2/ \partial y_1$ because of symmetry (see \cite{witt1990investigation} and \cite{schneider1984amplification}). 

The derivatives can be written as
\be \label{eqn:zetaone}
d\zeta=\frac{\partial \zeta}{\partial z} dz+\frac{\partial \zeta}{\partial \bar{z}} d\bar{z}=\left(\frac{\partial \zeta}{\partial z}+\frac{\partial \zeta}{\partial \bar{z}}\right) dy_1+i \left(\frac{\partial \zeta}{\partial z}-\frac{\partial \zeta}{\partial \bar{z}} \right) dy_2 \,
\ee
and
\bea \label{eqn:zetatwo}
d\zeta &=&du_1+i du_2=\frac{\partial u_1}{\partial y_1} dy_1+\frac{\partial u_1}{\partial y_2} dy_2+i \left(\frac{\partial u_2}{\partial y_1} dy_1+\frac{\partial u_2}{\partial y_2} dy_2 \right) \nonumber\\
&=& \left( \frac{\partial u_1}{\partial y_1}+i \frac{\partial u_2}{\partial y_1}\right) dy_1+\left( \frac{\partial u_1}{\partial y_2}+i \frac{\partial u_2}{\partial y_2}\right) dy_2 \,.
\eea
Using Eqs.~(\ref{eqn:zetaone},\ref{eqn:zetatwo}) we get
\blea
\frac{\partial \zeta}{\partial z}+\frac{\partial \zeta}{\partial \bar{z}}=\frac{\partial u_1}{\partial y_1}+i\frac{\partial u_2}{\partial y_1} \,,\\
\frac{\partial \zeta}{\partial z}-\frac{\partial \zeta}{\partial \bar{z}}=\frac{\partial u_2}{\partial y_2}-i\frac{\partial u_1}{\partial y_2} \,.
\elea
Solving for the derivatives of $\zeta$ and squaring gives
\blea
4 \left( \frac{\partial \zeta}{\partial z}\right)^2&=&\left(\frac{\partial u_1}{\partial y_1}+\frac{\partial u_2}{\partial y_2}\right)^2 \,, \\
4 \frac{\partial \zeta}{\partial \bar{z}} \overline{\frac{\partial \zeta}{\partial \bar{z}}}&=&\left(\frac{\partial u_1}{\partial y_1}-\frac{\partial u_2}{\partial y_2}\right)^2+4 \left(\frac{\partial u_2}{\partial y_1}\right)^2 \,.
\elea
Subtracting the previous equations gives the expression for the determinant of the Jacobian as given in Eq.~(\ref{eqn:jac}), so we can write
\be \label{eqn:jaczeta}
\det{J}=\left( \frac{\partial \zeta}{\partial z}\right)^2-\frac{\partial \zeta}{\partial \bar{z}} \overline{\frac{\partial \zeta}{\partial \bar{z}}} \,.
\ee
If we don't have continuous mass density, like in the case of lenses that can be approximated by point masses $\partial \zeta/ \partial z=1$ (see \cite{witt1990investigation}) so Eq.~(\ref{eqn:jaczeta}) becomes
\be \label{eqn:jfinal}
\det{J}=1-\frac{\partial \zeta}{\partial \bar{z}} \overline{\frac{\partial \zeta}{\partial \bar{z}}} \,.
\ee

To find the total magnification we simply add the absolute values of the magnifications of all the images
\be \label{eqn:magnsum}
A \equiv \sum_j |A_j| \,.
\ee
The places where the magnification diverges or $\det{J}=0$ form closed curves. These curves are called critical curves on the image plane and caustics on the source's plane. 

\section{Triple lens systems}
\label{sec:triple}

\subsection{Parametrization}

\begin{figure}[h]
	\centering
	\includegraphics[height=6cm]{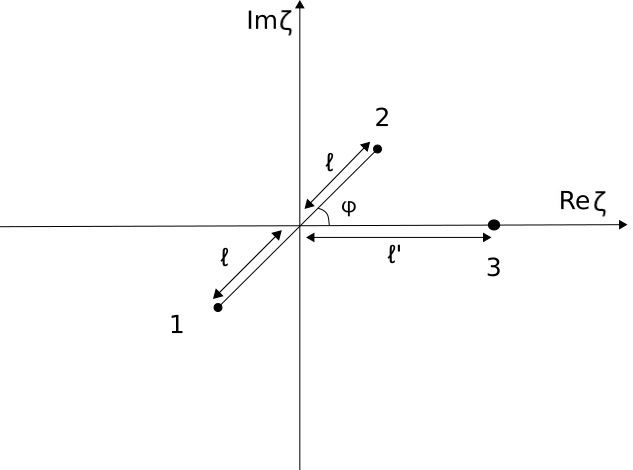}
	\caption{The figure shows the choice of parameters for the three lens system. The beginning of the axes is the middle between masses $1$ and $2$ while the position of mass $3$ is on the real axis. The angle $\phi$ is the angle between the line that connects masses $1$ and $2$ and the positive real axis. }
	\label{fig:threelens}
\end{figure}

For three lenses the $N$ lens Eq.~(\ref{eqn:lenscomplex}) becomes
\be \label{eqn:lensthree}
\zeta=z-\frac{\epsilon_1}{\bar{z}-\bar{z}_1}-\frac{\epsilon_2}{\bar{z}-\bar{z}_2}-\frac{\epsilon_3}{\bar{z}-\bar{z}_3} \,.
\ee
As in the case of a double lens system we can simplify the equation and reduce the number of parameters. First we notice that $\epsilon_1+\epsilon_2+\epsilon_3=1$ so we can write $\epsilon_3=1-\epsilon_1-\epsilon_2$. Then we can set the beginning of the axes in the middle between the two lenses ($1$ and $2$). So $z_2=-z_1$ and we can place $z_3 \equiv \ell'$ on the real axis. Finally we will write the $z_1$ in its polar form so $z_1=\ell e^{i \phi}$, where $\ell \equiv |z_1|$. So the lens equation (\ref{eqn:lensthree}) becomes
\be \label{eqn:lensthreepar}
\zeta=z-\frac{\epsilon_1}{\bar{z}-\ell e^{-i\phi}}-\frac{\epsilon_2}{\bar{z}+\ell e^{-i\phi}}-\frac{1-\epsilon_1-\epsilon_2}{\bar{z}-\ell'} \,.
\ee
So the system can be completely described by five parameters: $\epsilon_1, \epsilon_2, \ell, \ell', \phi$. Fig.~\ref{fig:threelens} represents the configuration.

\subsection{Method}
\label{sub:method}

In the single lens case we can easily solve the lens equation, find the positions and then calculate the magnification. Something similar could in principle be done in the $N$ lens case; solve the lens equation \eqref{eqn:lenscomplex} for $z$ and get $j$ solutions $z_j$ for the images positions then use the location of the images to calculate the magnification of each using Eq.~\eqref{eqn:magnsum}. However, this method presents significant difficulties even in the double lens case. 

If we examine the triple lens equation \eqref{eqn:lensthree}, we notice that it involves both $z$ and $\bar{z}$. To remove the $\bar{z}$ dependence we take the conjugate of Eq.~(\ref{eqn:lensthreepar}) which gives us the conjugate of $z$ 
\be \label{eqn:lensthreecon}
\bar{z}=\bar{\zeta}+\frac{\epsilon_1}{z-\ell e^{i\phi}}+\frac{\epsilon_2}{z+\ell e^{i\phi}}+\frac{1-\epsilon_1-\epsilon_2}{z-\ell'} \,.
\ee
Replacing in Eq.~(\ref{eqn:lensthreepar}) and clearing the denominators gives a tenth order polynomial in $z$
\be \label{eqn:polyone}
P_1(z,\zeta)=\sum_{i=1}^{10} c_i(\zeta) z^i=0 \,.
\ee
Rhie \cite{rhie2002cumbersome} presents a way to calculate the coefficients analytically \footnote{However, the manuscript seems to contain errors.}.  

Locating the image positions from the equation and then finding the magnification for each of them is not only a difficult task but also unnecessary; we do not need the image positions but the total magnification.  So instead of solving this tenth order equation we find the magnification directly using the resultant method. This method was first developed by Mao and Witt \cite{witt1995minimum} for the double lens case. 

First we evaluate
\be
\frac{\partial{\zeta}}{\partial{\bar{z}}}=\frac{\epsilon_1}{(\bar{z}-\ell e^{-i\phi})^2}+\frac{\epsilon_2}{(\bar{z}+\ell e^{-i\phi})^2}+\frac{1-\epsilon_1-\epsilon_2}{(\bar{z}-\ell')^2} \,,
\ee
using Eq.~\eqref{eqn:lensthreepar}. Then from Eq.~\eqref{eqn:magnindi} and \eqref{eqn:jfinal} we have the magnification of each image $A_j$ in terms of $z$, $\zeta$ and $\bar{z}$. Using \eqref{eqn:lensthreecon} we eliminate $\bar{z}$ and after clearing the denominators we end up with an 18th order polynomial in $z$
\be
P_2(A_j,z,\zeta)=0 \,.
\ee
Now we have two polynomials in $z$ with coefficients that are functions of the source position $\zeta$ and the magnification $A_j$. As our goal is to find the $A_j$'s in terms of $\zeta$, we eliminate $z$ by taking the resultant of $P_1$ and $P_2$. This was done using Mathematica and in particular using the ``subresultant'' method which was proven to be the most efficient. The result is a polynomial in $A_j$ with coefficients that are functions of $\zeta$
\be
P_{\text{res}}(A_j,\zeta)=0 \,.
\ee
The roots of the polynomial are the individual magnifications of each image, and summing their absolute values \eqref{eqn:magnsum} gives the total magnification. 

Setting $\zeta=x+iy$ we can then plot the magnification at each point for different values of the source position creating a magnification map. In the double lens case one can find these roots and proceed to plot the magnification. However, in the triple lens case that proved impossible. Instead, we used the Mathematica random number generator to select pairs of $(x,y)$ and proceeded to evaluate the magnification for each individual point. A computation of about 20,000 points which creates a detailed magnification map takes about $45$ min processor computation time with a 2.3 GHz Quad-Core Intel Core i7 processor.

\begin{figure}
	\centering
	\includegraphics[height=6cm]{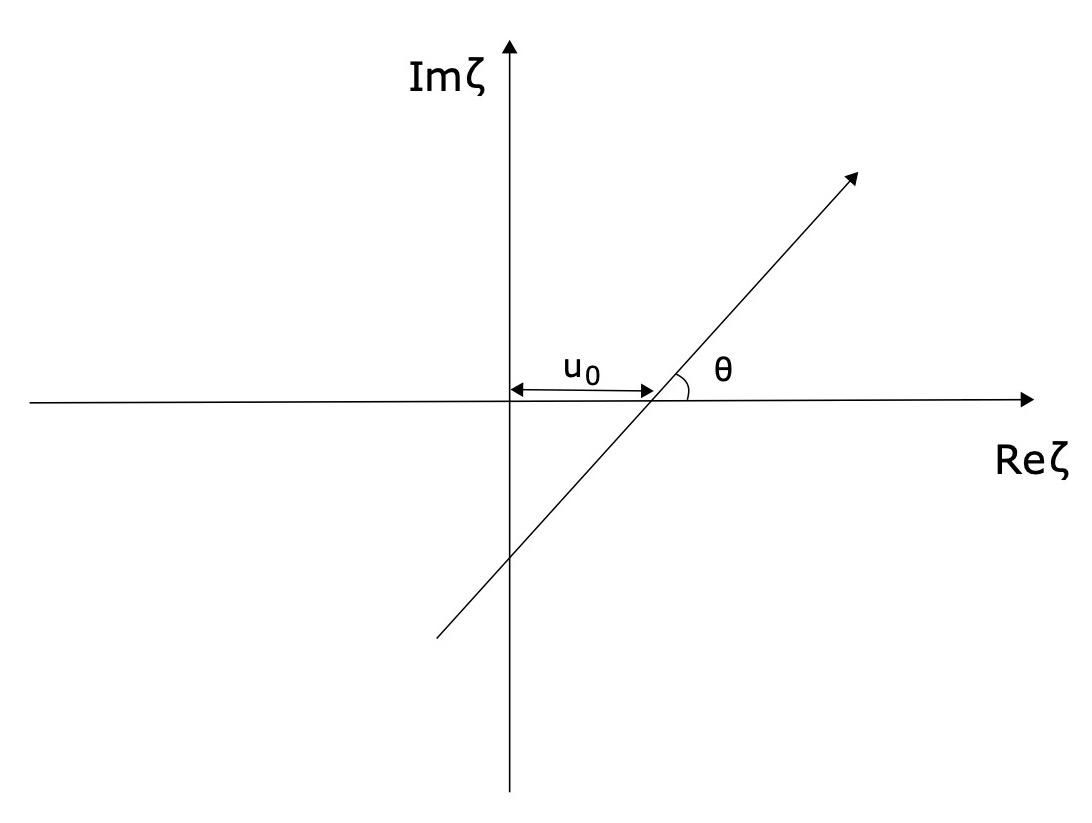}
	\caption{The figure shows the choice of parameters for the source trajectory.}
	\label{fig:sourcetraj}
\end{figure}

In order to generate light curves we first parameterize the source trajectory. Here, $\theta$ is the angle that the source trajectory makes with the $x$ axis of the source plane. We define $u_0$ as the distance on the $x$ axis of the source trajectory from the origin. The parametrization is shown in figure \ref{fig:sourcetraj}. Note that this is a slightly different definition of $u_0$ than the one used in literature. Then we have
\be
y=x\tan{\theta}-u_0 \,.
\ee
As in the magnification maps we produce points $\{x,|A|\}$ using a random number generator for the $x$ values. 

\section{Exploration of parameters}
\label{sec:parameters}

In this section we use the method described in Sec.~\ref{sub:method} to derive magnification maps and light curves for a variety of circumbinary systems. As an assumption, the lenses $1$ and $2$ will have larger mass than lens $3$ to model a system with two stars and a planet. We proceed to examine the effects of the of the mass ratios, the distance between the stars, the distance of the planet and the angle $\phi$. 

We start with setting our base parameters. In each section one of these parameters will vary and the rest will be set to the following values unless otherwise stated. The mass of the stars is set equal to one another and the mass of the planet $1/1000$ of that. The stars distance is $\ell=0.25$, the planet distance $\ell'=1$ and the angle of the system $\phi=\pi/4$. The planet distance is taken longer than the stars distance to explore possible microlensing configurations for planets further from their stars where other methods fail. The trajectory of the source is characterized by $\theta=\pi/4$ and $u_0=0.1$. 

\newpage

\subsection{Mass of planet}

\begin{wrapfigure}{L}{0.7\linewidth}
	\centering
	\includegraphics[scale=0.6]{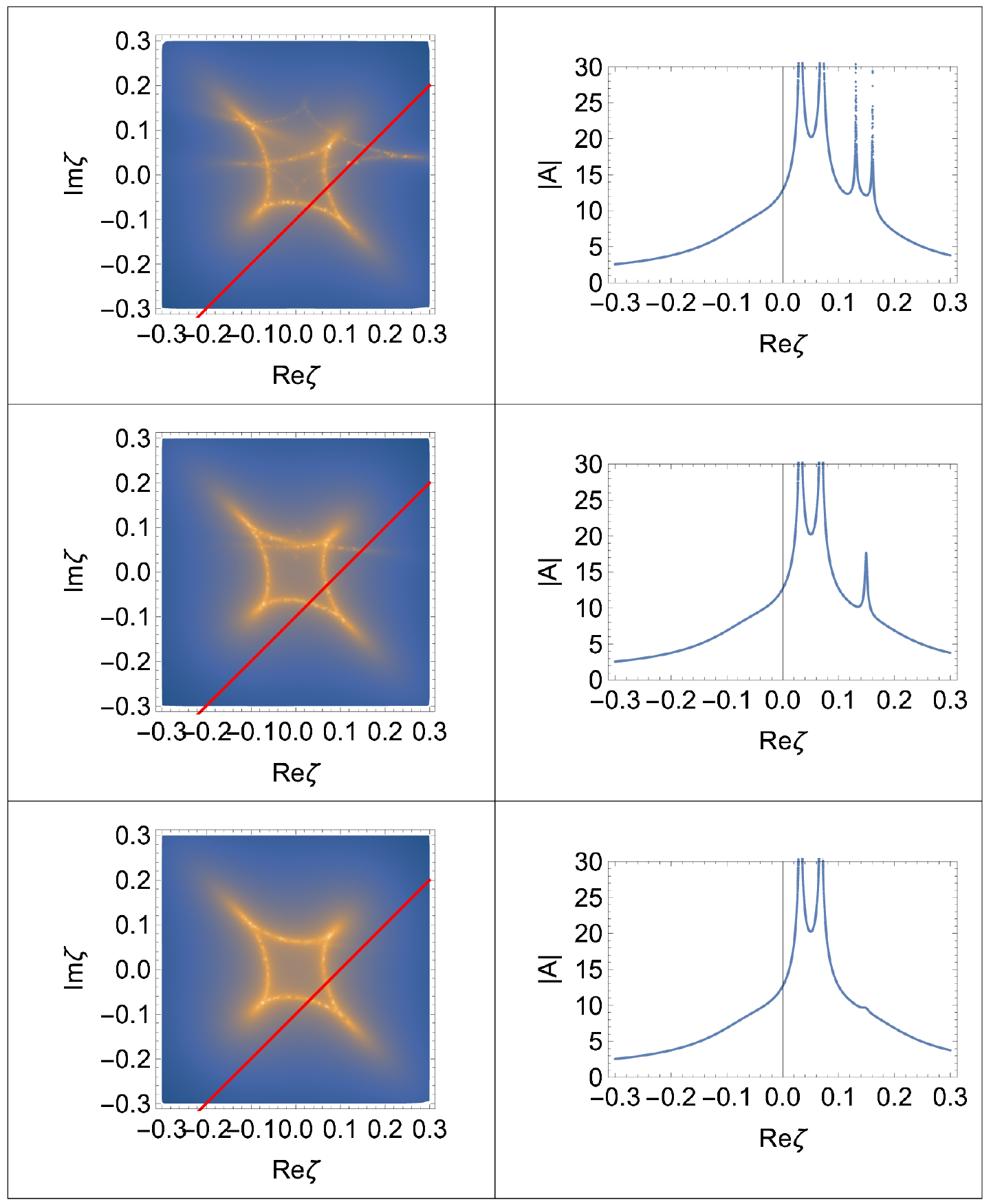}
	\caption{Magnification maps and light curves for different star to planet ratios: $1/100$, $1/1000$, $1/10000$ (from top to bottom). The density plots on the left show the logarithmic absolute magnification on the source plane. The curves on the right show the light curve produced for the given trajectory of the source shown in red on the left.}
	\label{fig:massplanet}
\end{wrapfigure}

First we explore the effect of the mass of the planet compared to the stars in a circumbinary system. We plot three different planetary masses expressed as the ratio of the mass of the star to the planet: $1/100$, $1/1000$ and $1/10000$. The first could correspond to a system of two main sequence Sun-size stars and a brown dwarf, an object that has a mass between those of the largest gas giant planets and the smallest stars. The second represents a giant gas planet the size of Jupiter and the third a smaller gas planet the size of Neptune. 

The magnification maps and light curves are given in Fig.~\ref{fig:massplanet}. For the brown dwarf case the caustic of the system is complex and characteristic of a triple lens system. The peak corresponding to the presence of the brown dwarf is double and very visible. In the case of the Jupiter-size planet the planetary caustic is a perturbation to the main caustic of the stars. However, looking at the light curve, the planet produces a very clear peak easily detectable. In the last case, the planetary caustic is barely visible and the light curve of the stars is only slightly perturbed.

\subsection{Distance of planet}

\begin{figure}
	\centering
	\includegraphics[scale=0.6]{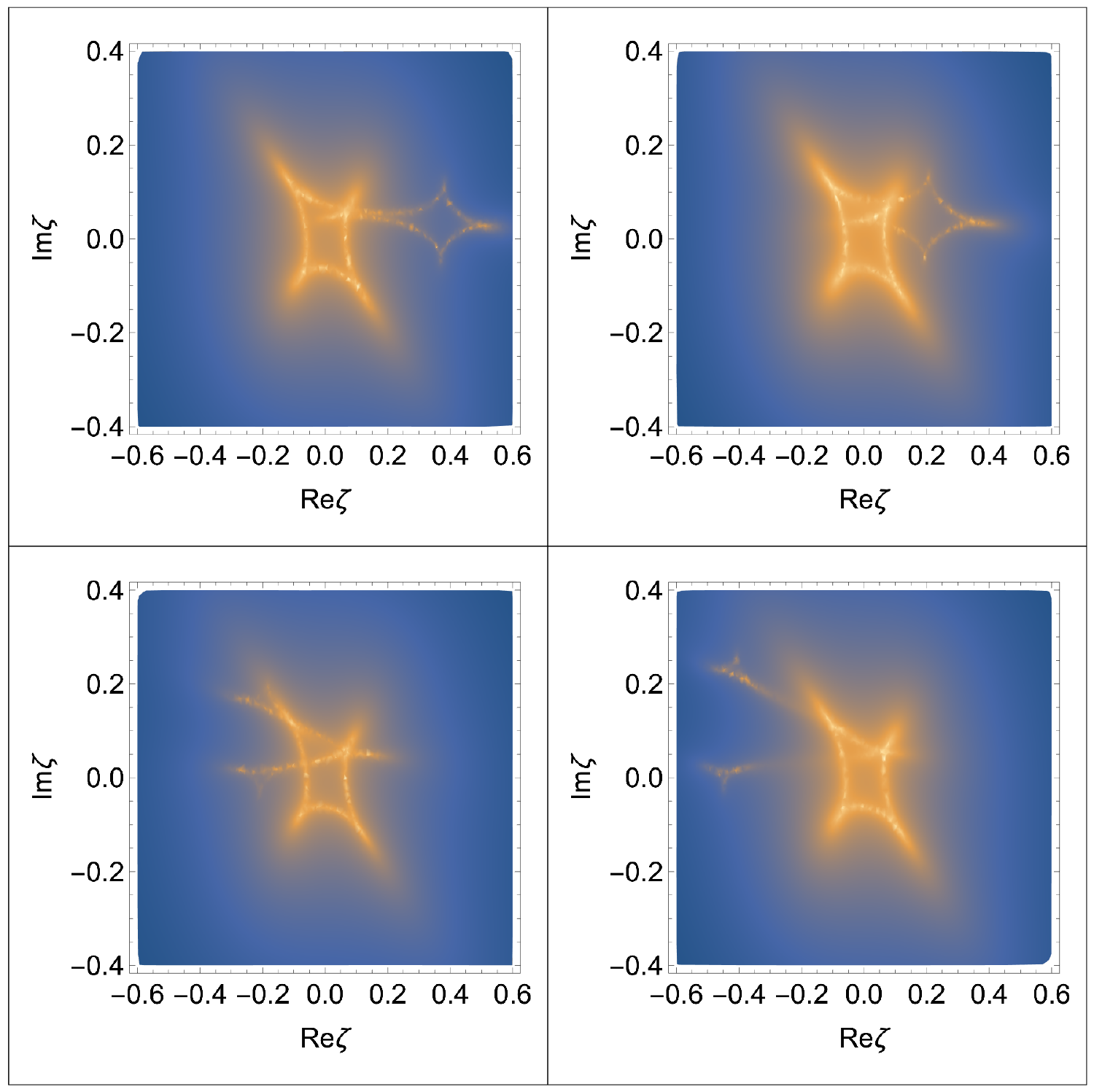}
	\caption{Magnification maps for different planet distances: $\ell'=1.2$, $1.1$, $0.9$ and $0.8$ (from top left clockwise). The top panels show wide systems while the bottom ones close.}
	\label{fig:distanceplanet}
\end{figure}

Varying the distance of the planet has a significant effect in the shape of the planetary caustics. In the case of a star and planet system the three topologies of the caustic are close ($<1$), resonant ($=1$) and wide ($>1$) \cite{Gaudi:2010nk}. The same categories can be applied to a circumbinary system with the relevant distance being the one of the planet from the center of two stars $\ell'$. These topologies have been studied at the level of caustics \cite{luhn2016caustic} but here we present exact magnification maps. In Fig.~\ref{fig:distanceplanet} we show four magnification maps with different planet distances using the base parameters but planet mass ratio $1/100$. The choice of the large mass ratio was in order have the ability to note the change of topology between the wide and close systems.  It is interesting to note that the planet is detectable more easily in the more wide configuration (top left $\ell'=1.2$), a case where utilizing other methods such as the transit method is difficult.

\subsection{Distance of stars}

 The distance between the two stars can change both the shape of the planetary caustic and the light curve of the system. When the two stars are close, the planetary caustic is a result of both stars as they are similar distances to the planet. 
 
\begin{figure}
	\centering
	\includegraphics[scale=0.6]{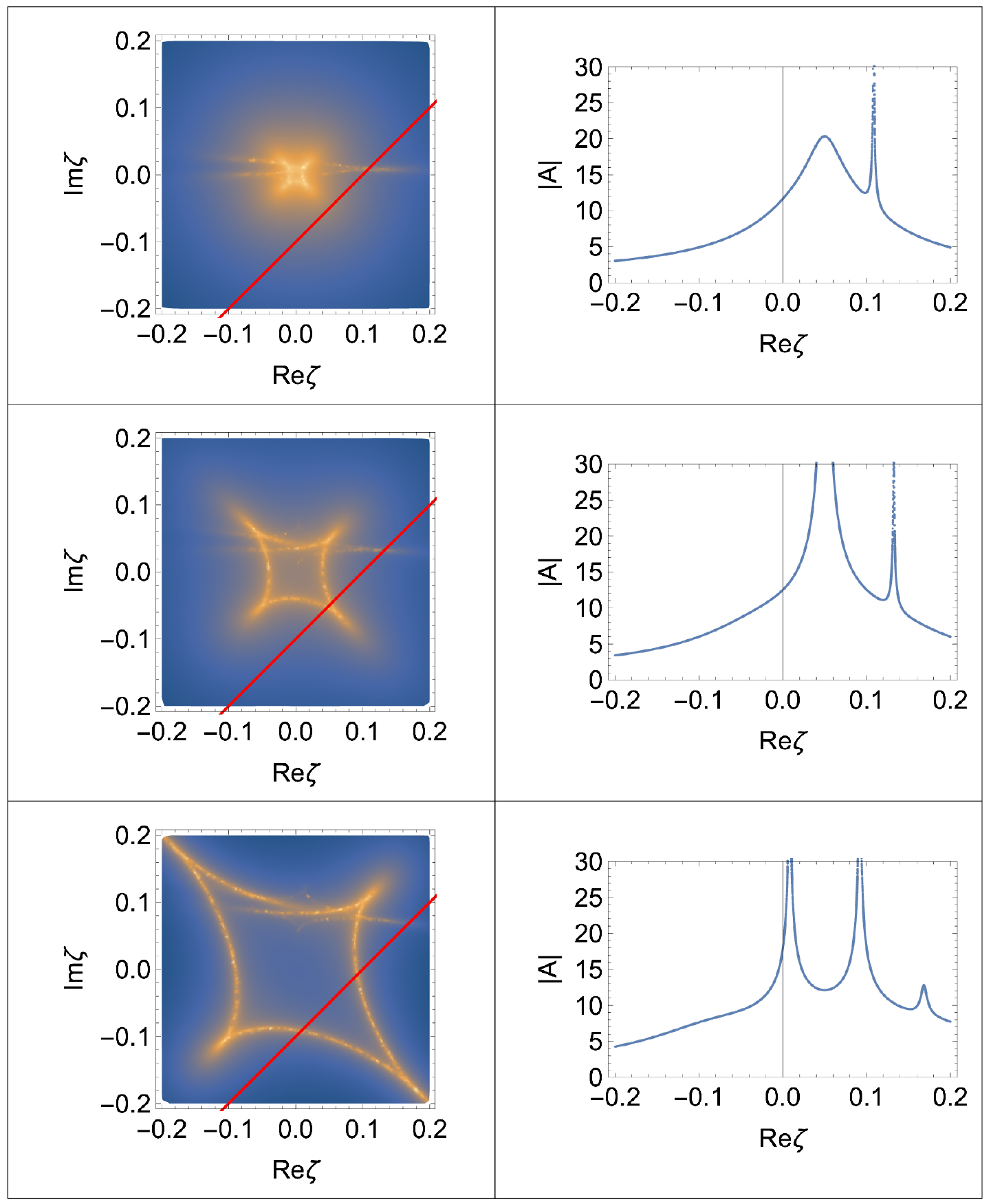}
	\caption{Magnification maps and light curves for different star separation: $\ell=0.1$, $0.2$, $0.3$ (from top to bottom). The density plots on the left show the logarithmic absolute magnification on the source plane. The curves on the right show the light curve produced for the given trajectory of the source shown in red on the left.}
	\label{fig:starsdistancesmall}
\end{figure}

 In Fig.~\ref{fig:starsdistancesmall} we examine magnification maps and light curves for short star distances and mass ratio $1/1000$. On the top panel the stars are so close ($\ell=0.1$) that the corresponding light curve is similar to a single star and a planet. In such a system it would be difficult to detect both stars and the planet as the caustics of the stars are dominant for sources passing through the center of the system. The second panel shows a case where the stars are further apart ($\ell=0.2$) and dependent on the trajectory of the source, both stars could be detected or not. The third case corresponds to a larger distance ($\ell=0.3$) where both stars could be detected for most source trajectories. It is interesting to observe that the planetary caustic is less prominent in this case.

\begin{figure}
	\centering
	\includegraphics[scale=0.6]{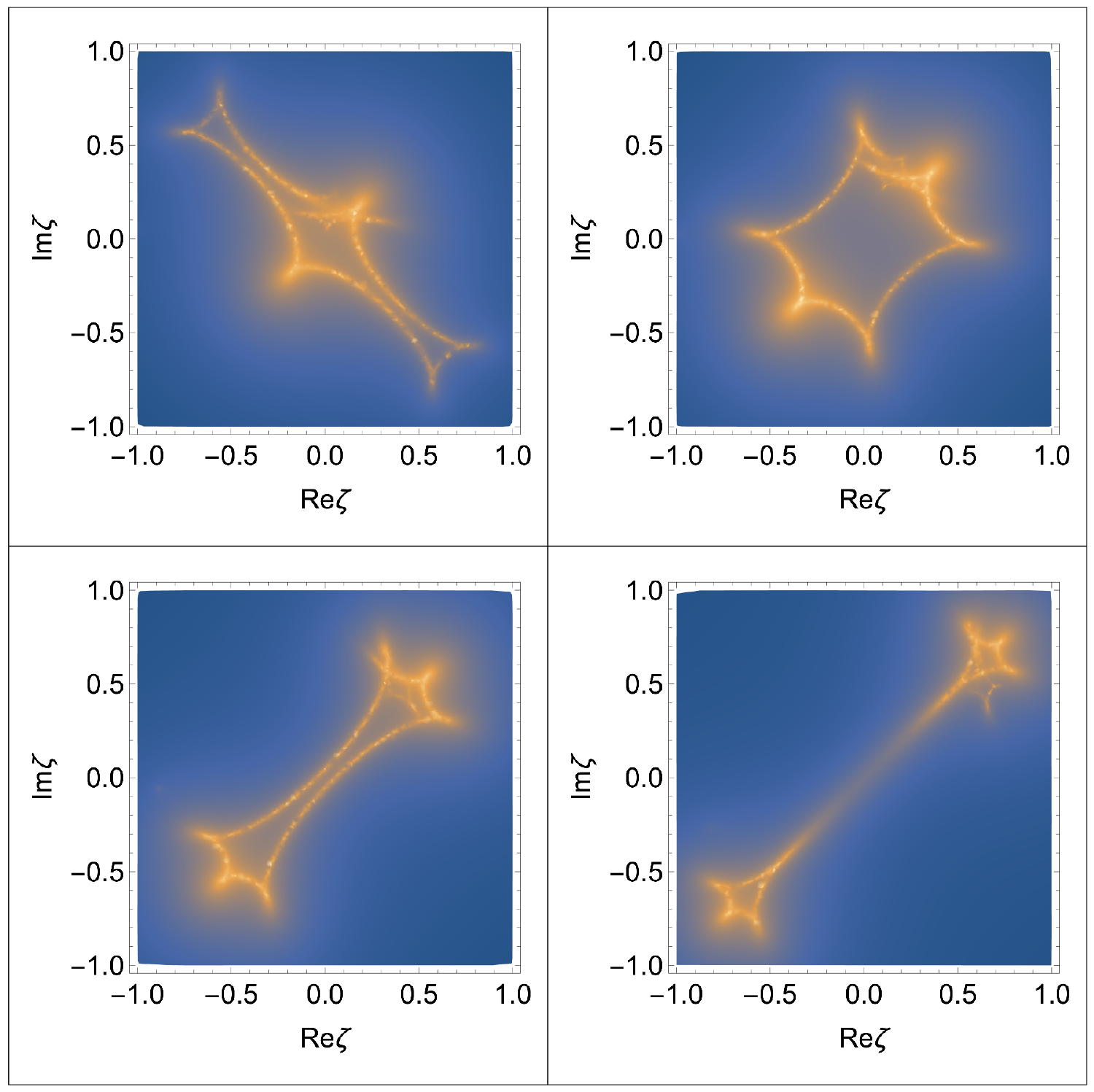}
	\caption{Magnification maps for different star distances: $\ell=0.4$, $0.65$, $0.9$ and $1.15$ (from top left clockwise). The top panels show the change in geometry for $\ell \gtrapprox 0.5$ while thew bottom ones the weakening of the caustic connecting the two stars for $\ell \gtrapprox 1$.}
	\label{fig:starsdistancelarge}
\end{figure}

 In Fig.~\ref{fig:starsdistancelarge} we examine magnification maps for larger star distances. We use mass ratio $1/100$ so that the planetary caustic is visible. Here, with increasing star distance the geometry increasingly resembles that of a single star and planet with the stars caustics disconnecting. The geometry of the caustic between the two stars changes drastically from $\ell=0.4$ (top left panel) to $\ell=0.65$ (top right panel). At the bottom left ($\ell=0.9$) and especially bottom right panel ($\ell=1.15$) we see the stars' caustic separating. We should note that even though the caustic connecting the two stars has visibly less magnification, the shape of it is still that of a double lens. The planetary caustic is moving along with the star that it is closer to, even though the position of the planet does not change.

\newpage

\subsection{Angle of the system}

\begin{wrapfigure}{R}{0.7\linewidth}
		\centering
		\includegraphics[scale=0.6]{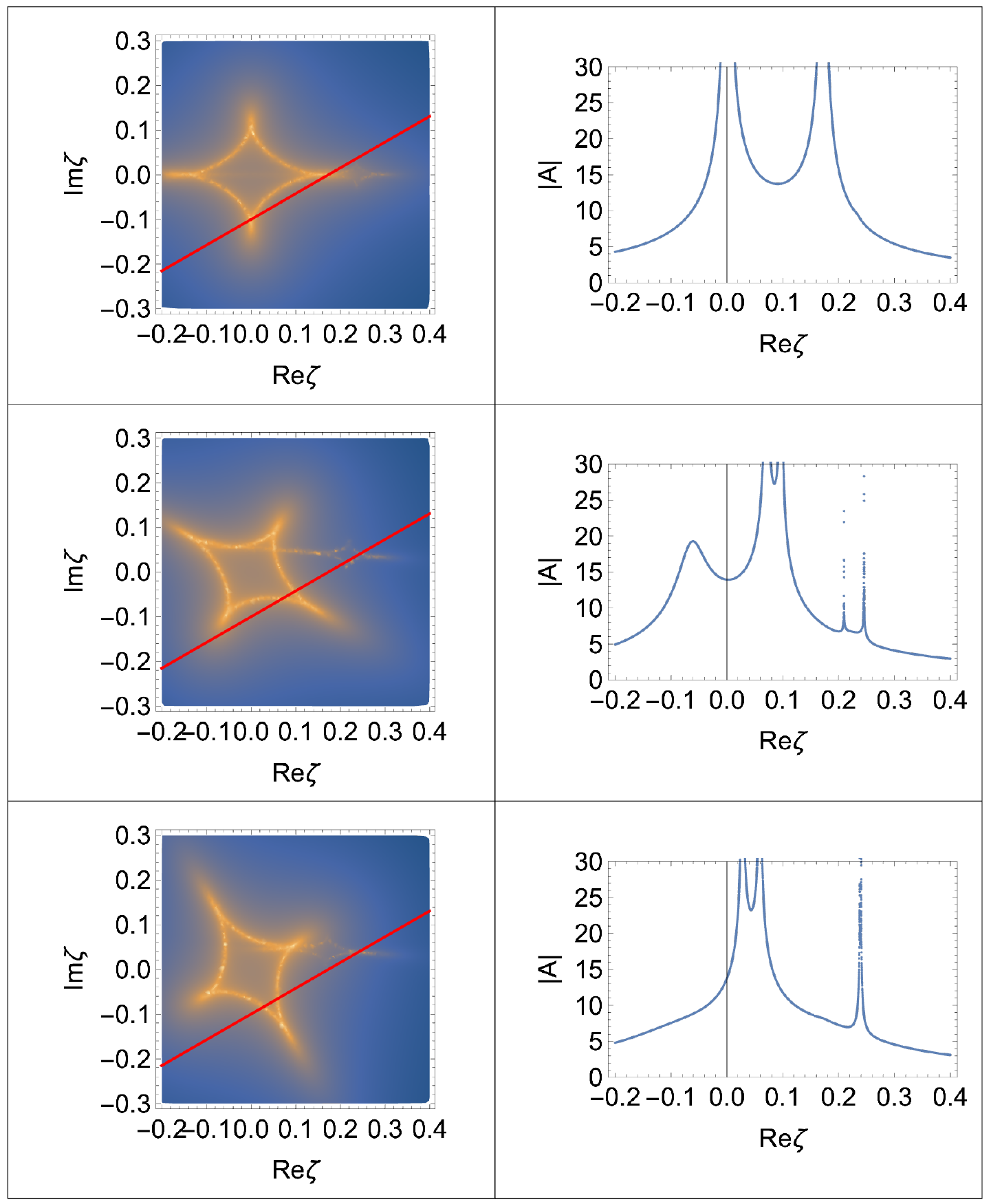}
		\caption{Magnification maps and light curves for different angles of the system: $\phi=\pi/2, \pi/3, \pi/6$ (from top to bottom). The density plots on the left show the logarithmic absolute magnification on the source plane. The curves on the right show the light curve produced for the given trajectory of the source shown in red on the left.}
		\label{fig:angle}
	\end{wrapfigure}

Next we vary the angle $\phi$ of the system. It is obvious that different angles lead to very different geometries of the caustics and different detectability of the planet in the system. In the plots shown in Fig.~\ref{fig:angle} we use $\ell'=1.1$ so the planetary caustic is distinct. 

As we notice in Fig.~\ref{fig:angle} the planetary caustic moves further up the vertical axis and closer to the center as the angle of the system gets smaller. In terms of detectability of the planet, the most difficult case is that of the planet being on or close to the symmetry axis of the star system. For most source trajectories apart from the horizontal, ones the planet can remain undetectable.

\begin{figure}[h]
	\centering
	\includegraphics[scale=0.6]{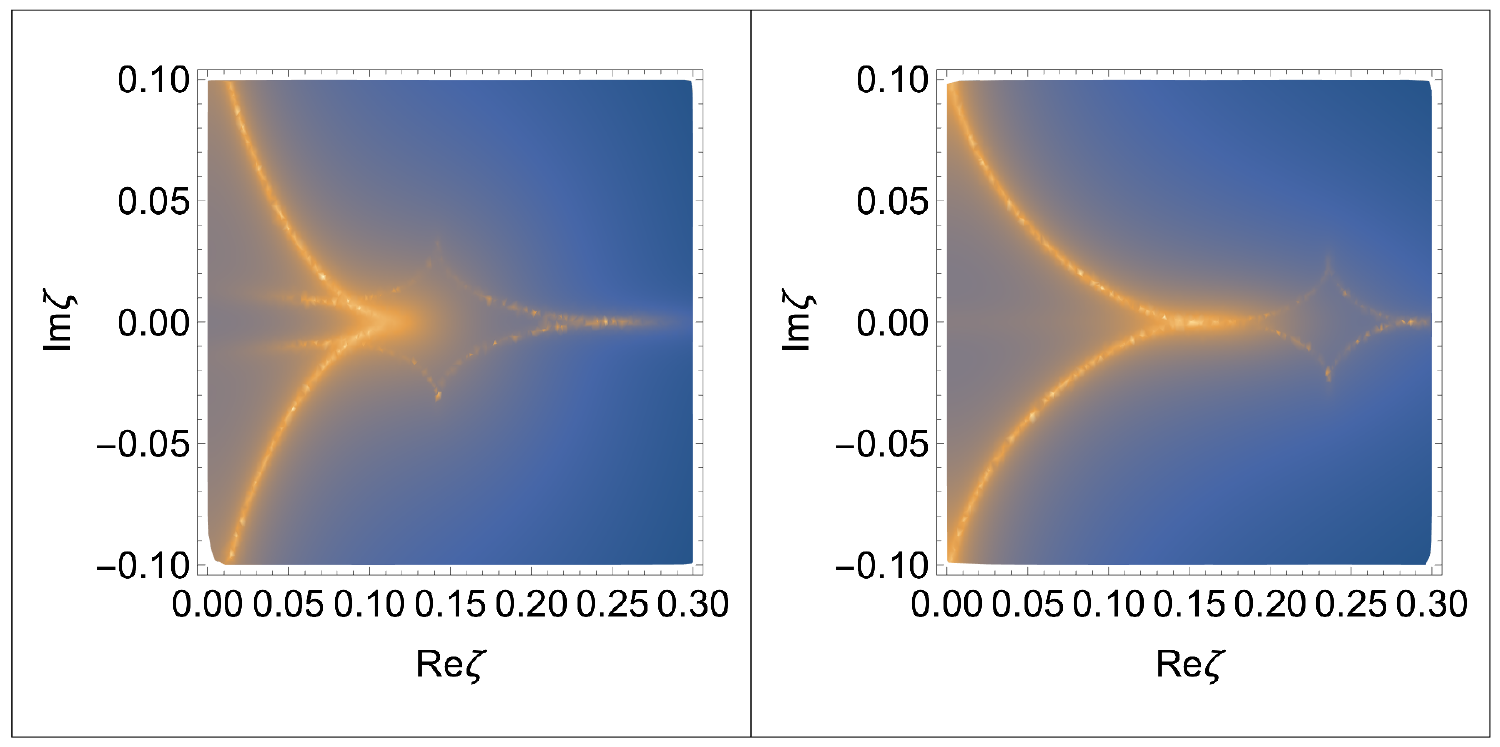}
	\caption{Magnification maps of the $\phi=\pi/2$ (left) and $\phi=0$ (right) angles of the system.}
	\label{fig:zoom}
\end{figure}

An interesting comparison is that of $\phi=0$ and $\phi=\pi/2$ shown in Fig.~\ref{fig:zoom}. Here the symmetry of the two systems is the same, making the planet difficult to detect. However, there are significant differences in the shape of the planetary caustic. For $\phi=0$ the caustic is much closer and it shows the beginnings of a close geometry. For $\phi=\phi/2$ we have a clear wide geometry with the planetary caustic being smaller and further away.

\subsection{Mass of stars}

In this subsection we vary the mass of the two stars to produce magnification plots for systems with two different masses. We are especially interested in the cases where the system transitions from that of two stars and a planet to one star and a planet. In practice this may be difficult to distinguish observationally, as often both models fit well on the light curve.

\begin{figure}[h]
	\centering
	\includegraphics[scale=0.6]{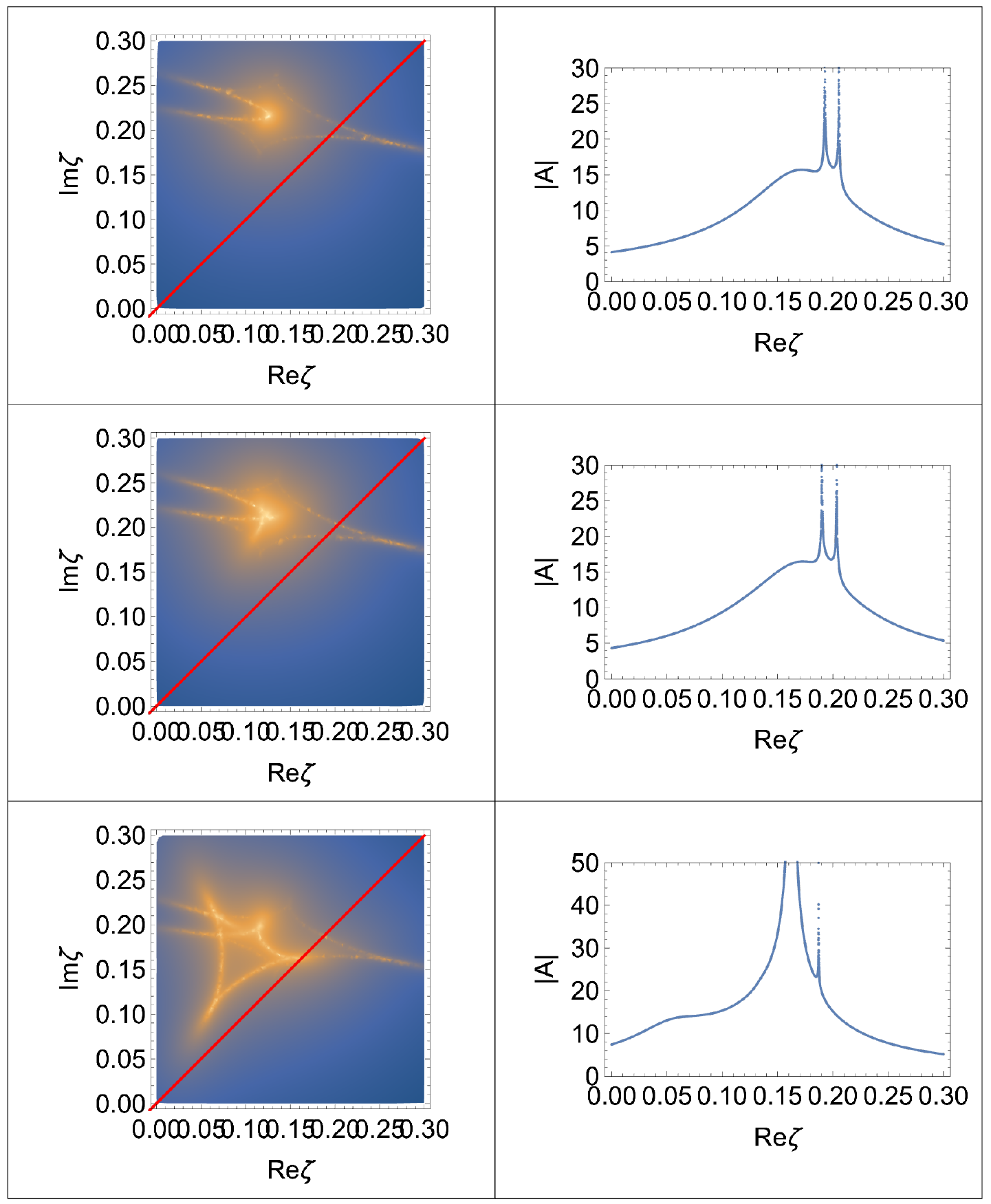}
	\caption{Magnification maps and light curves with $\epsilon_1/\epsilon_2=1/1000$, $1/100$ and $1/10$ (top to bottom). here $\phi=\pi/3$ and $u_0=0$. }
	\label{fig:starsmass}
\end{figure}

To examine this transition we picked three cases:  two stars of unequal mass and a planet, a star, a brown dwarf and a planet, and a star and two planets of equal mass. Fig.~\ref{fig:starsmass} shows the three systems, the ratio of the largest to the smallest mass is always $1/1000$ while the second mass $\epsilon_2$ takes the values $1/1000$, $1/100$ and $1/10$ of $\epsilon_1$ (top to bottom). The top corresponds to a system with two planets with one of them close to the star. The light curve is indistinguishable of a system with a star and one planet as the second planet is too close to the star. In the second case, the geometry changes as the planet (or brown dwarf) is significantly larger. However, this is not necessarily detectable from a light curve far enough from the star. The third case represents the  geometry of two stars with one of them larger than the other. 
	
\section{Example: the OGLE-2007-BLG-349L(AB)c system}
\label{sec:example}

The first circumbinary system discovered using microlensing (and the only one thus far) was OGLE-2007-BLG-349L(AB)c in 2016 \cite{bennett2016first}. The light curve from the system fits two classes of models: 2-planet models and circumbinary planet models. Here we use the best fit data for both the two planet and the circumbinary models (Table 2 of \cite{bennett2016first}). 

In order to match the data, we should note that in the conventions of \cite{bennett2016first}, the mass $\epsilon_1$ corresponds to our mass $\epsilon_3$. Then the origin of the axes is the center of mass between masses $\epsilon_1$ and $\epsilon_2$ and not the middle of the distance. Thus we need to make the substitution $\ell \to \epsilon_1 \ell/ \epsilon_2$. Table~\cite{tab:OGLE} translates the data to match our conventions. 

	\begin{table}[h]
\centering \begin{tabular}{|c|c|c|}
		\hline
		 & Circumbinary & 2-planet \\
		 \hline
	$\epsilon_1$ & $0.46479$ & $8.5025\times 10^{-6}$  \\
	$\epsilon_2$ & $0.53487$ &$0.999615$\\
	$\ell$ & $0.0106502$ & $0.950452$ \\
	$\ell'$ & $-0.81468$ & $-0.79607$ \\
	$\phi$(rads) & $0.36989$ & $-3.07611$\\
	\hline
\end{tabular}
\caption{Parameters of the system for magnification maps.}
	\end{table}

There are several differences between our method and the one used by \cite{bennett2016first}. We solve the system exactly and produce complete magnification maps that show more than the caustics. However, we do not take into account the orbital motion of the system and the size of the source. So the magnification shown in the light curves is larger than the one in the \cite{bennett2016first} model. These parameters could be incorporated in future work but are beyond the scope of this paper.

\begin{figure}[h]
	\centering
	\includegraphics[scale=0.6]{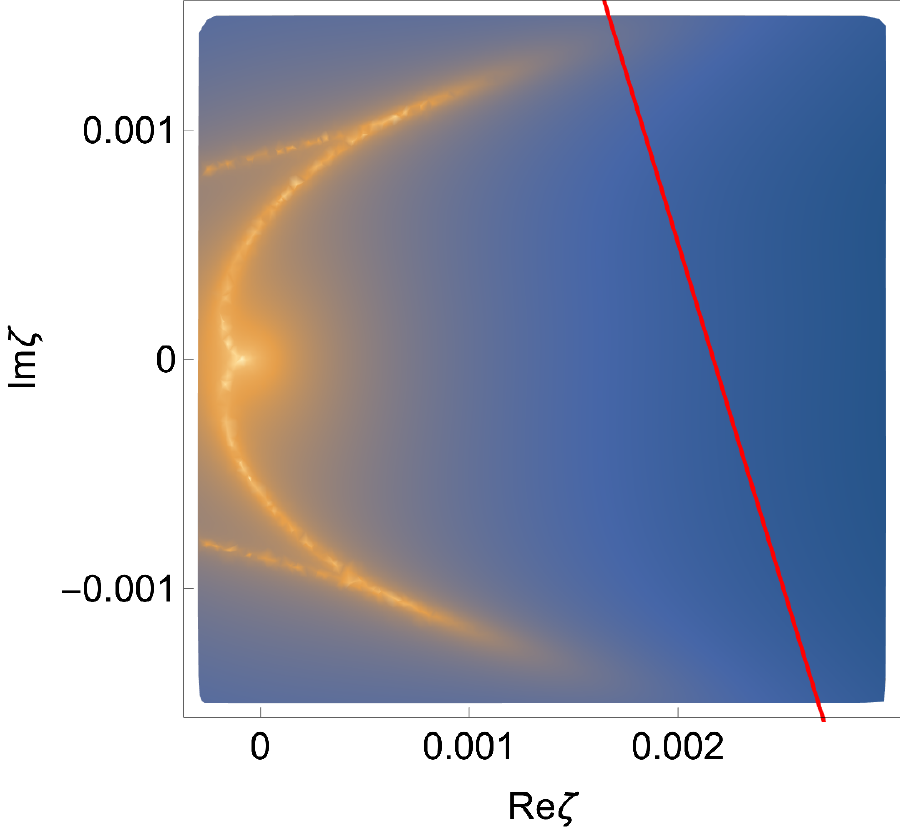}
	\includegraphics[scale=0.6]{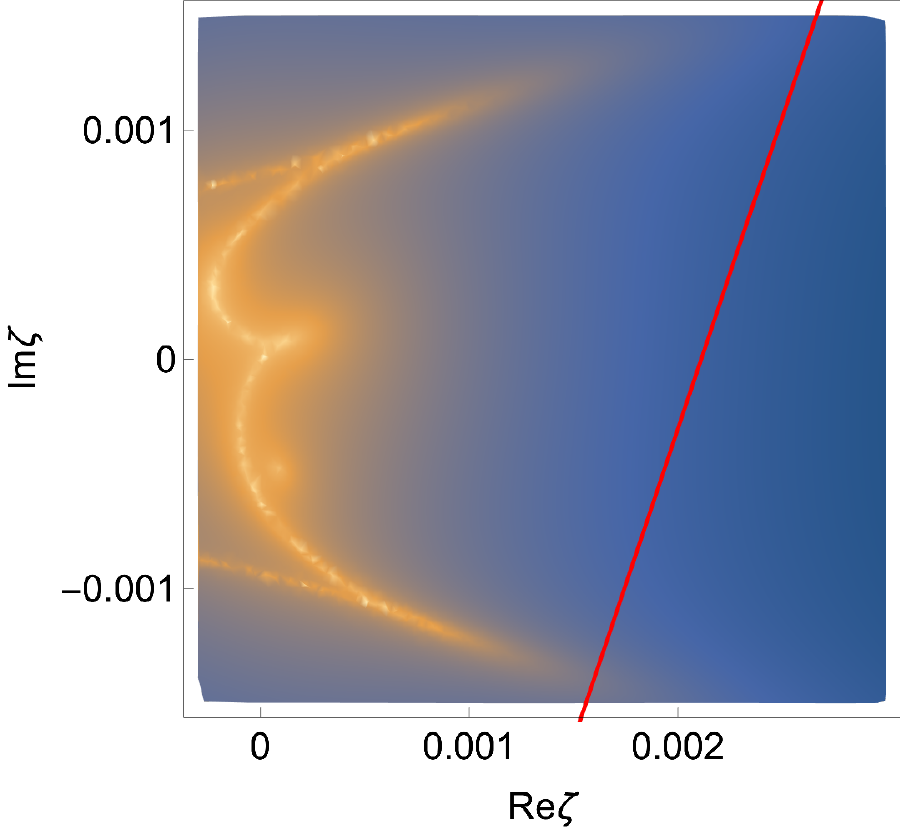}
	\caption{Magnification maps for the two probable geometries: the two planet (left) and circumbinary (right).}
	\label{fig:magnOGLE}
\end{figure}

First we look at the magnification maps that correspond to the caustics of Fig.~3 of \cite{bennett2016first}. Looking at Fig.~\ref{fig:magnOGLE} the magnification maps look relatively similar. The main difference is the $y$-axis symmetry which is broken in the case of the circumbinary system. 

	\begin{figure}[h]
	\centering
\includegraphics[scale=0.8]{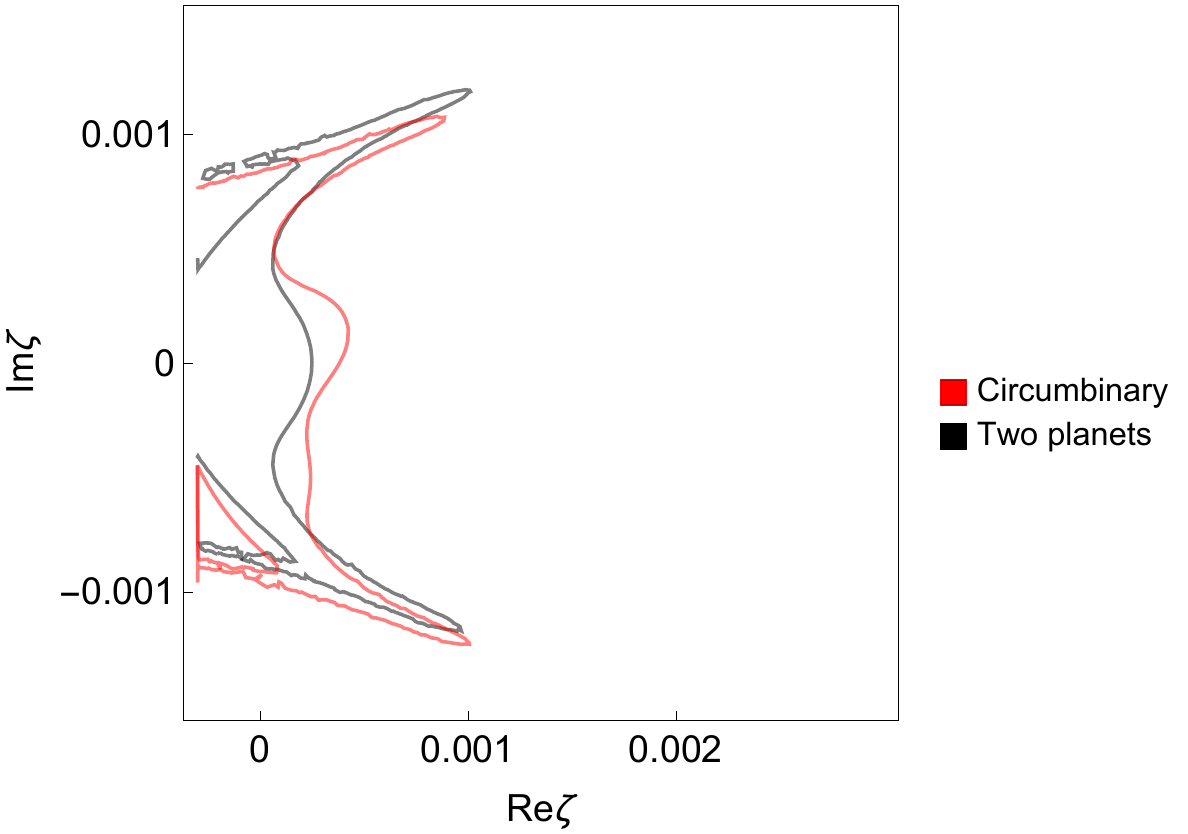}
	\caption{Comparison of the $\log{|A|}>8$ for the two planet and the circumbinary geometries.}
	\label{fig:compmagnOGLE}
\end{figure}

A comparison of the contour with $\log{|A|}>8$ in Fig.~\ref{fig:compmagnOGLE} shows the differences between the two geometries in more detail.  In particular, the circumbinary case has larger magnification near the center. The symmetry differences between the two cases are also illustrated. 

\begin{figure}[h]
	\centering
	\includegraphics[scale=0.9]{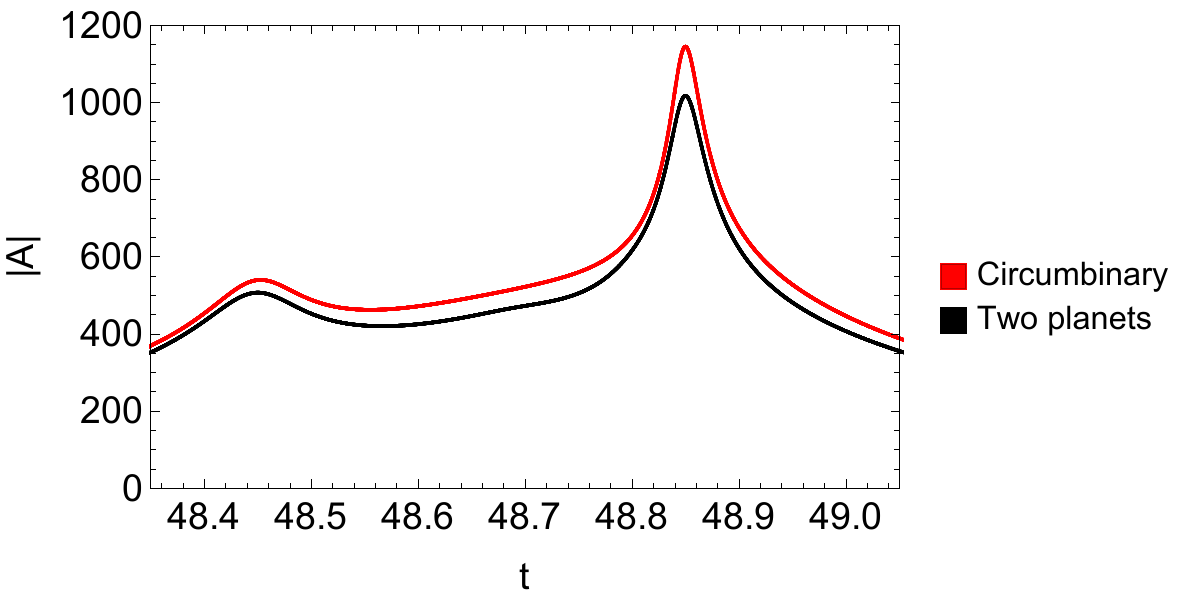}
	\caption{Light curves for the two geometries using the two different source trajectories. The form of the curves is almost identical. The time $t$ is in $HJD'$ following \cite{bennett2016first}.}
	\label{fig:OLGElightshort}
\end{figure}

Turning to the light curves, we use the parameters of \cite{bennett2016first} for the source trajectory in each case. While the magnification maps have differences, with the right source trajectory the two geometries have almost identical light curves (see Fig.~\ref{fig:OLGElightshort}). The difference in maximum  magnification can be attributed to the finite source versus point source as used in this analysis. 

\section{Conclusions}
\label{sec:conclusions}

In this work we explored the parameter space of circumbinary systems in exoplanetary microlensing using an exact method of solution for the three lens equation. The effect of all five parameters was considered and we produced both exact magnification maps and light curves. The magnification maps reveal a richer structure than the caustics usually studied. 

There are several directions in which this work could be extended. Including orbital motion of the system is an important one. Creating animations of magnification maps can show the change of the geometry of the system similar to what was done in \cite{luhn2016caustic}. It would also be very interesting to combine the orbital motion with the relative motion of the source and generate the corresponding light curves. In real microlensing events, the orbital motion can play a role if its time scale is comparable to the transit time of the source. 

Including a finite source with uniform or non-uniform surface brightness is also of interest. Important theoretical work has been done on that topic (e.g. \cite{witt1994can}) so those results could be incorporated. 

A longer term goal would be to extend the method to include more lenses. New techniques are likely needed here as the complexity of the system with even one more lens would make the numerical results significantly slower. The much larger parameter space can also pose a problem which could be circumvented by imposing certain additional symmetries. 

\vspace{0.5in}

\begin{footnotesize}
	\noindent{\bf Acknowledgments}
E-AK and ET would like to thank Hal Haggard for collaboration in the earlier stages of this work. ET was supported by the Physics Program of Bard College. BG was supported by the Research Associates program of the College of the Holy Cross. E-AK is supported by the ERC Consolidator Grant QUANTIVIOL.
\end{footnotesize}

\bibliographystyle{utphys}
	\bibliography{circumbinary}

\end{document}